\RequirePackage{lineno}
\documentclass[prl,twocolumn,amsmath,amssymb,showpacs,superscriptaddress,nofootinbib,table]{revtex4-1}
\usepackage[table]{xcolor}
\usepackage[dvipdfm,CJKbookmarks=true,unicode,colorlinks,linkcolor=blue,anchorcolor=blue,citecolor=blue,pdfborder={0 0 0}]{hyperref} 
\usepackage{graphicx,epsfig,fancyhdr,amsfonts,amssymb,subfigure,psfrag}
\usepackage{dcolumn} 
\usepackage{bm}
\makeatletter
    \def\CT@@do@color{%
      \global\let\CT@do@color\relax
            \@tempdima\wd\z@
            \advance\@tempdima\@tempdimb
            \advance\@tempdima\@tempdimc
    \advance\@tempdimb\tabcolsep
    \advance\@tempdimc\tabcolsep
    \advance\@tempdima2\tabcolsep
            \kern-\@tempdimb
            \leaders\vrule
                    \hskip\@tempdima\@plus  1fill
            \kern-\@tempdimc
            \hskip-\wd\z@ \@plus -1fill }
\makeatother

\usepackage{multirow}
\usepackage{booktabs}
\usepackage[none]{hyphenat}

\newcommand{\eq}[1]{\begin{equation} #1 \end{equation}}

\newcommand*{\jpsi}{J/\psi}
\newcommand*{\pip}{\pi^{+}}
\newcommand*{\pim}{\pi^{-}}
\newcommand*{\etap}{\eta^{\prime}}

\newcommand*{\piz}{\pi^{0}}
\newcommand*{\gev}{\text{GeV}/c^{2}}
\newcommand*{\mev}{\text{MeV}/c^{2}}
\newcommand*{\logl}{\log\mathcal{L}}
\newcommand*{\ppbar}{p\bar{p}}

\newcommand*{\gpp}{g^{2}_{p\bar{p}}/g^{2}_{0}}

\newcommand*{\costheta}{\text{cos}~\theta}
\renewcommand*{\to}{\rightarrow}

\begin{document}
\preprint{}
\sloppy
\title{\boldmath Observation of an anomalous line shape of the $\eta^{\prime}\pi^{+}\pi^{-}$ mass spectrum near the $p\bar{p}$ mass threshold in $J/\psi\rightarrow\gamma\eta^{\prime}\pi^{+}\pi^{-}$}

\author{\small
  M.~Ablikim$^{1}$, M.~N.~Achasov$^{9,e}$,
  S.~Ahmed$^{14}$, X.~C.~Ai$^{1}$, O.~Albayrak$^{5}$,
  M.~Albrecht$^{4}$, D.~J.~Ambrose$^{44}$, A.~Amoroso$^{49A,49C}$,
  F.~F.~An$^{1}$, Q.~An$^{46,a}$
% ...
  \\
  \vspace{0.2cm}
  (BESIII Collaboration)\\
  \vspace{0.2cm} {\it
    $^{1}$ Institute of High Energy Physics, Beijing 100049, People's Republic of China\\
% ...
$^{4}$ Bochum Ruhr-University, D-44780 Bochum, Germany\\
    $^{5}$ Carnegie Mellon University, Pittsburgh, Pennsylvania 15213, USA\\
% ...
$^{9}$ G.I. Budker Institute of Nuclear Physics SB RAS (BINP), Novosibirsk 630090, Russia\\
% ...
$^{14}$ Helmholtz Institute Mainz, Johann-Joachim-Becher-Weg 45, D-55099 Mainz, Germany\\
% ...
$^{44}$ University of Rochester, Rochester, New York 14627, USA\\
% ...
$^{46}$ University of Science and Technology of China, Hefei 230026, People's Republic of China\\
% ...
$^{49}$ (A)University of Turin, I-10125, Turin, Italy; (B)University of Eastern Piedmont, I-15121, Alessandria, Italy; (C)INFN, I-10125, Turin, Italy\\
% ...
$^{a}$ Also at State Key Laboratory of Particle Detection and Electronics, Beijing 100049, Hefei 230026, People's Republic of China\\
% ...
$^{e}$ Also at the Novosibirsk State University, Novosibirsk, 630090, Russia\\
% ...
  }\vspace{0.4cm}}


\begin{abstract}
  Using $1.09\times10^{9}$ $J/\psi$ events collected by the BESIII experiment in 2012,
  we study the $J/\psi\rightarrow\gamma\eta^{\prime}\pi^{+}\pi^{-}$ process and
  observe a significant abrupt change in the slope of the $\eta^{\prime}\pi^{+}\pi^{-}$ invariant mass distribution at the
  proton-antiproton ($p\bar{p}$) mass threshold.
  We use two models to characterize the $\eta^{\prime}\pi^{+}\pi^{-}$ line shape
  around $1.85~\text{GeV}/c^{2}$: one which explicitly incorporates the opening of a
  decay threshold in the mass spectrum (Flatt\'{e} formula), and another
  which is the coherent sum of two resonant amplitudes.
  Both fits show almost equally good agreement with data,
  and suggest the existence of either a broad state around $1.85~\text{GeV}/c^{2}$ with strong couplings to $p\bar{p}$ final states
  or a narrow state just below the $p\bar{p}$ mass threshold.
  Although we cannot distinguish between the fits, either one supports the existence of a $p\bar{p}$ molecule-like
  state or bound state with greater than $7\sigma$ significance.
\end{abstract}

\maketitle

The state $X(1835)$ was first observed by the BESII experiment as a peak
in the $\etap\pip\pim$ invariant mass distribution in
$\jpsi\to\gamma\etap\pip\pim$ decays~\cite{x1835_bes2}. This observation was later
confirmed by BESIII studies of the same process~\cite{x1835_bes3}
with mass and width measured to be $M=1836.5\pm3^{+5.6}_{-2.1}~\mev$ and $\Gamma=190\pm9^{+38}_{-36}~\mev$;
the $X(1835)$ was also observed in the $\eta K^{0}_{S} K^{0}_{S}$ channel in $\jpsi\to\gamma\eta K^{0}_{S}K^{0}_{S}$
decays, where its spin-parity was determined to be $J^{P}=0^{-}$ by a partial
wave analysis (PWA)~\cite{x1835_qiny}.
An anomalously strong enhancement at the proton-antiproton ($\ppbar$)
mass threshold, dubbed $X(\ppbar)$,
was first observed by BESII in $\jpsi\to\gamma\ppbar$ decays~\cite{xpp_bes2};
this observation was confirmed by BESIII~\cite{xpp_bes3} and CLEO~\cite{xpp_cleo}.
This enhancement structure was subsequently determined to have spin-parity
$J^{P}=0^{-}$ by BESIII~\cite{xpp_bes3pwa}.
Among the various theoretical
interpretations on the nature of the $X(1835)$ and
$X(\ppbar)$~\cite{int1,int2,int3,int4,int5}, a particularly intriguing one
suggests that the two structures originate from a $\ppbar$ bound
state~\cite{ppint1,ppint2,ppint3,ppint4,ppint5}.  If the $X(1835)$ is really
a $\ppbar$ bound state, it should have a strong coupling to $0^{-}$
$\ppbar$ systems, in which case the line shape of $X(1835)$ at the $\ppbar$
mass threshold would be affected by the opening of the $X(1835)\to\ppbar$ decay
mode. A study of the $\etap\pip\pim$ line shape of $X(1835)$ with high
statistical precision therefore provides valuable information that helps clarify
the nature of the $X(1835)$ and $X(\ppbar)$.

In this Letter, we report the observation of a significant abrupt change in slope
of the $X(1835)\to\etap\pip\pim$ line shape at the $\ppbar$ mass threshold
in a sample of $\jpsi\to\gamma\etap\pip\pim$ events collected in the BESIII
detector at the BEPCII $e^+ e^-$ storage ring. The $\etap$ is reconstructed in its
two major decay modes: $\etap\to\gamma\pip\pim$ and
$\etap\to\eta\pip\pim,~\eta\to\gamma\gamma$. The data sample used in
this analysis contains a total of $1.09\times10^{9}$ $\jpsi$
decay events~\cite{jpsi-number} accumulated by the BESIII experiment in 2012.

The BESIII detector~\cite{bes3} is a magnetic spectrometer operating at BEPCII~\cite{bepc2},
a double-ring $e^{+}e^{-}$ collider with center of mass energies between 2.0 and 4.6 GeV.
The cylindrical core of the BESIII detector consists of a helium-based main drift chamber
(MDC), a plastic scintillator time-of-flight system (TOF), and a CsI(Tl) electromagnetic calorimeter
(EMC) that are all enclosed in a superconducting solenoidal magnet providing a 0.9 T magnetic field.
The solenoid is supported by an octagonal flux-return yoke with resistive plate counter muon
identifier modules interleaved with steel.  The acceptance of charged particles and photons
is 93\% of the 4$\pi$ solid angle. The charged-particle momentum resolution at 1 GeV/c is 0.5\%;
the EMC measures 1~GeV photons with an energy resolution of 2.5\% (5\%) in the barrel (end cap)
regions.  A GEANT4-based~\cite{geant4} Monte Carlo (MC) simulation software package is used to
optimize the event selection criteria, estimate backgrounds, and determine the detection efficiency.
The KKMC~\cite{kkmc} generator is used to simulate $\jpsi$ production.

Charged tracks are reconstructed using hits in the MDC.  The point of closest approach of each
charged track to the $e^{+}e^{-}$ interaction point is required to be within 20~cm in the
beam direction and 2~cm in the plane perpendicular to the beam direction. The reconstructed polar
angle between the charged-track and beam direction is restricted to $|\costheta|<0.93$.
The TOF and energy loss ($dE/dx$) information are combined to form particle identification confidence levels for the $\pi$, $K$, and $p$ hypotheses;
each track is assigned to the particle type that corresponds to the hypothesis with the highest confidence level.
Photon candidates are selected from showers in the EMC with energy deposited in the EMC barrel
($|\costheta|<0.8$) or end-cap regions ($0.86<|\costheta|<0.92$) to be greater than 100 MeV.
EMC cluster timing requirements are used to suppress electronic noise and energy deposits that are
unrelated to the event.

\begin{figure*}[htbp]
  \centering
  \includegraphics[width=0.95\textwidth]{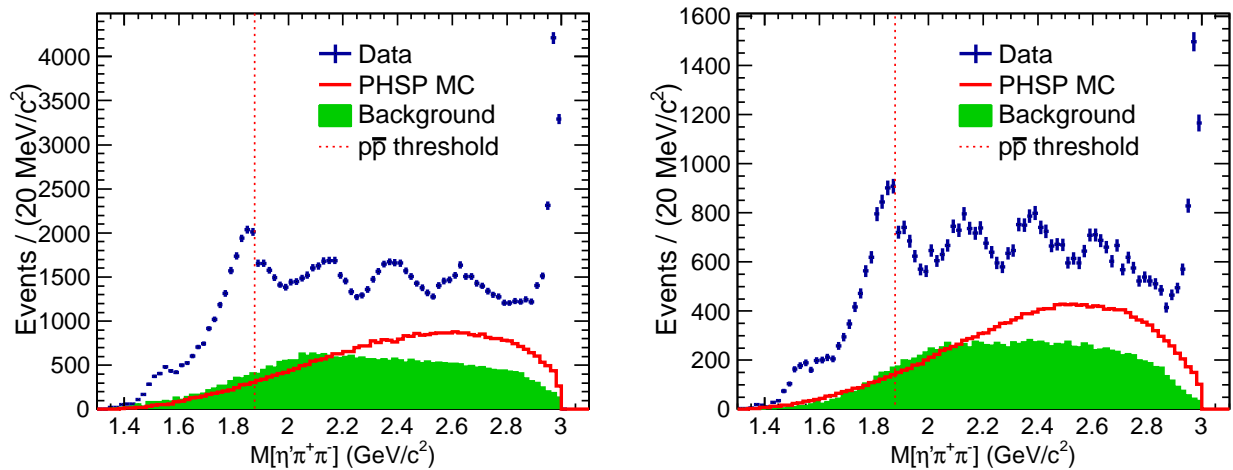}
  \caption{The $\etap\pip\pim$ invariant mass spectra after the application of all selection criteria.
  The plot on the left side shows the spectrum for events with the $\etap\to\gamma\pip\pim$ channel;
  that on the right shows the spectrum for the $\etap\to\eta(\to\gamma\gamma)\pip\pim$ channel.
  In both plots, the dots with error bars are data, the shaded histograms are the background,
  the solid histograms are phase space (PHSP) MC events of $\jpsi\to\gamma\etap\pip\pim$ (arbitrary normalization),
  the dotted vertical line shows the position of $\ppbar$ mass threshold.}
  \label{fig_mX}
\end{figure*}

To select events from $\jpsi\to\gamma\etap\pip\pim$, four charged tracks with zero net charge are required.
Among the four tracks, at least three of them should be positively identified as pions.
At least two photon candidates are required for the $\etap\to\gamma\pip\pim$ mode;
three photon candidates are required for the $\etap\to\eta\pip\pim,~\eta\to\gamma\gamma$ mode.

For the $\jpsi\to\gamma\etap(\to\gamma\pip\pim)\pip\pim$ channel, a four-constraint (4C) kinematic fit
which imposes energy and momentum conservation is performed to the $\gamma\gamma\pip\pim\pip\pim$ hypothesis;
the $\chi^{2}_{4C}$ of the kinematic fit is required to be smaller than 40.
If there are more than two photon candidates, the two-photon combination with the smallest $\chi^{2}_{4C}$ value
is retained.
Events with: $|M_{\gamma\gamma}-m_{\piz}|<40~\mev$; $|M_{\gamma\gamma}-m_{\eta}|<30~\mev$;
$720~\mev<M_{\gamma\gamma}<820~\mev$; or $400~\mev<M_{\gamma\pip\pim}<563~\mev$ are rejected to suppress
background events from: $\jpsi\to\piz\pip\pim\pip\pim$; $\jpsi\to\eta\pip\pim\pip\pim$;
$\jpsi\to\omega(\to\gamma\piz)\pip\pim\pip\pim$; and
$\jpsi\to\gamma\eta(\to\gamma\pip\pim,~\piz\pip\pim)\pip\pim$, respectively.
Finally, the $\etap$ is reconstructed by a $\gamma\pip\pim$ combination that satisfies both
$|M_{\pip\pim}-m_{\rho^{0}}|<200~\mev$ and $|M_{\gamma\pip\pim}-m_{\etap}|<15~\mev$. If two or more combinations
pass these two criteria, the one with the smallest $|M_{\gamma\pip\pim}-m_{\etap}|$ is chosen.

For the $\jpsi\to\gamma\etap(\to\eta\pip\pim, \eta\to\gamma\gamma)\pip\pim$ channel, a 4C kinematic fit
to the $\gamma\gamma\gamma\pip\pim\pip\pim$ hypothesis is performed; events with $\chi^{2}_{4C}<40$ are accepted.
If there are more than three photon candidates, the three that minimize $\chi^{2}_{4C}$ are retained.
To suppress backgrounds from $\piz\to\gamma\gamma$,
events in which any one of the three two-photon pairings satisfies $|M_{\gamma\gamma}-m_{\piz}|<40~\mev$ are rejected.
The $\eta$ is reconstructed by the two photons that best satisfy $|M_{\gamma\gamma}-m_{\eta}|<30~\mev$.
A five-constraint (5C) kinematic fit, energy-momentum conservation with an additional constraint on
the $\eta\to\gamma\gamma$ invariant mass to be equal to $m_{\eta}$, is performed, and $\chi^{2}_{5C}<40$ is required.
Then the $\etap$ candidate is formed from the $\eta\pip\pim$ combination that best satisfies
$|M_{\eta\pip\pim}-m_{\etap}|<10~\mev$.

The $\etap\pip\pim$  invariant mass spectra of the surviving events are shown in Fig.~\ref{fig_mX},
where peaks corresponding to the $X(1835)$, $X(2120)$, $X(2370)$, $\eta_{c}$~\cite{x1835_bes3}, and a
structure near $2.6~\gev$ that has not been seen before are evident for both $\etap$ decays.
Thanks to the high statistical precision, an abrupt change in slope of the $X(1835)$ line shape at the
$\ppbar$ mass threshold is evident in both event samples.


An inclusive sample of $10^{9}$ $\jpsi$ decays events that are generated according to the Lund-Charm
model~\cite{lund-charm} and Particle Data Group (PDG)~\cite{pdg} decay tables, is used to study
potential background processes.  These include events with no real $\etap$s in the final state (non-$\etap$)
and those from $\jpsi\to\piz\etap\pip\pim$. We use $\etap$ mass sideband events to estimate the non-$\etap$
background contribution to the $\etap\pip\pim$ invariant mass distribution.
For the $\jpsi\to\piz\etap\pip\pim$ background, we use a one-dimensional data-driven method that
first selects $\jpsi\to\piz\etap\pip\pim$ from the data to determine the shape of their contribution
to the selected $\etap\pip\pim$ mass spectrum and re-weight this shape by the ratio of MC-determined efficiencies
for $\jpsi\to\gamma\etap\pip\pim$ and $\jpsi\to\piz\etap\pip\pim$ events;
the total weight after re-weighting is the estimated number of $\jpsi\to\piz\etap\pip\pim$ background events.
Our studies of background processes show
that neither the four peaks mentioned above nor the abrupt change in the line shape at $2m_p$ are caused by
background processes.

\begin{figure}[htbp]
  \centering
  \includegraphics[width=0.48\textwidth]{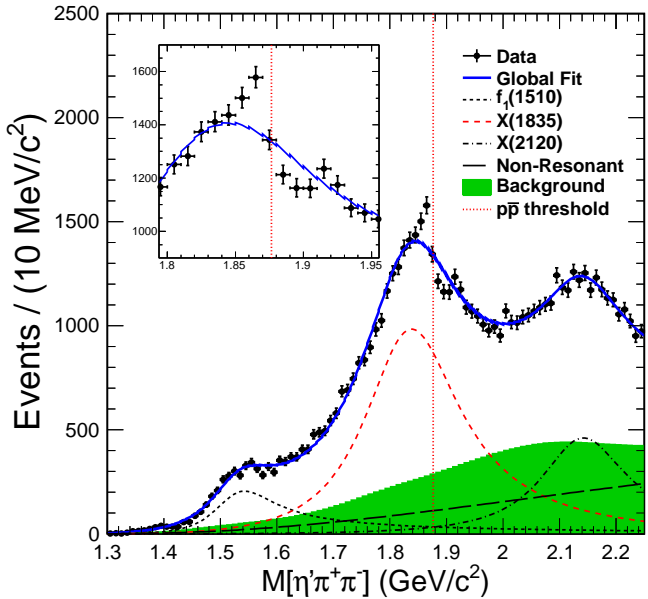}
  \caption{Fit results with simple Breit-Wigner formulae.
           The dashed dotted vertical line shows the position of $\ppbar$ mass threshold,
           the dots with error bars are data, the solid curves are total fit results, the dashed curves are the $X(1835)$,
           the short-dashed curves the $f_{1}(1510)$, the dash-dot curves the $X(2120)$, and
           the long-dashed curves are the non-resonant $\etap\pip\pim$ fit results;
           the shaded histograms are background events.
           The inset shows the data and the global fit between $1.8~\gev$ and $1.95~\gev$.}
  \label{fig_fit_bw}
\end{figure}

We perform simultaneous fits to the $\etap\pip\pim$ invariant mass distributions
between $1.3~\gev$ and $2.25~\gev$  for  both selected event samples with
the $f_{1}(1510)$, $X(1835)$ and $X(2120)$ peaks represented by
three efficiency-corrected Breit-Wigner functions convolved with a Gaussian function to account for the mass resolution,
where the Breit-Wigner masses and widths are free parameters.
The non-resonant $\etap\pip\pim$ contribution is obtained from Monte-Carlo simulation;
the non-$\etap$ and $\jpsi\to\piz\etap\pip\pim$ background contributions are obtained as discussed above.
For resonances and the non-resonant $\etap\pip\pim$ contribution,
the phase space for $\jpsi\to\gamma\etap\pip\pim$ is considered:
according to the $J^{P}$ of $f_{1}(1510)$ and $X(1835)$,
$\jpsi\to\gamma f_{1}(1510)$ and $\jpsi\to\gamma X(1835)$ are $S$-wave and $P$-wave processes, respectively;
all other processes are assumed to be $S$-wave processes.
Without explicit mention, all components are treated as incoherent contributions.
In the simultaneous fits, the masses and widths of resonances, as well as the branching fraction for $\jpsi$
radiative decays to $\etap\pip\pim$ final states (including resonances and non-resonant $\etap\pip\pim$)
are constrained to be the same for both $\etap$ decay channels. The fit results are shown in Fig.~\ref{fig_fit_bw},
where it is evident that using a simple Breit-Wigner function to describe the $X(1835)$ line shape fails near the
$\ppbar$ mass threshold.
The $\logl$ ($\mathcal{L}$ is the combined likelihood of simultaneous fits) of this fit is 630503.3.
Typically, there are two circumstances where an abrupt distortion of
a resonance's line shape shows up: a threshold effect caused by the opening of an additional decay mode;
or interference between two resonances.  We tried to fit the data for both of these possibilities.

In the first model, we assume the state around $1.85~\gev$ couples to $\ppbar$.
The line shape of $\etap\pip\pim$ above the $\ppbar$ threshold is therefore affected by the
opening of the $X(1835)\to\ppbar$ decay channel, similar to the distortion of the $f_0(980)\to\pip\pim$
line shape at the $K\bar{K}$ threshold.  To study this, the Flatt\'{e} formula~\cite{flatte} is used for the
$X(1835)$ line shape:
\eq{
  T=\frac{\sqrt{\rho_\text{out}}}{\mathcal{M}^{2}-s-i\sum_{k}g^{2}_{k}\rho_{k}}.
  \label{eq_flatte}
}
Here $T$ is the decay amplitude, $\rho_\text{out}$ is the phase space for $\jpsi\to\gamma\etap\pip\pim$,
$\mathcal{M}$ is a parameter with the dimension of mass,
$s$ is the square of the $\etap\pip\pim$ system's mass,
$\rho_{k}$ is the phase space for decay mode $k$, and $g^{2}_{k}$ is the corresponding coupling strength.
The term $\sum_{k}g^{2}_{k}\rho_{k}$ describes how the decay width varies with $s$. Approximately:
\eq{
  \sum_{k}g^{2}_{k}\rho_{k}\approx g^{2}_{0}\left(\rho_{0}+\frac{g^{2}_{p\bar{p}}}{g^{2}_{0}}\rho_{\ppbar}\right),
  \label{eq_partialwidth}
}
where $g^{2}_{0}$ is the sum of $g^{2}$ of all decay modes other than $X(1835)\to\ppbar$,
$\rho_{0}$ is the maximum two-body decay phase space volume~\cite{pdg} and
$\gpp$ is the ratio between the coupling strength to the $\ppbar$ channel and the sum of all other channels.

\begin{figure}[htbp]
  \centering
  \includegraphics[width=0.48\textwidth]{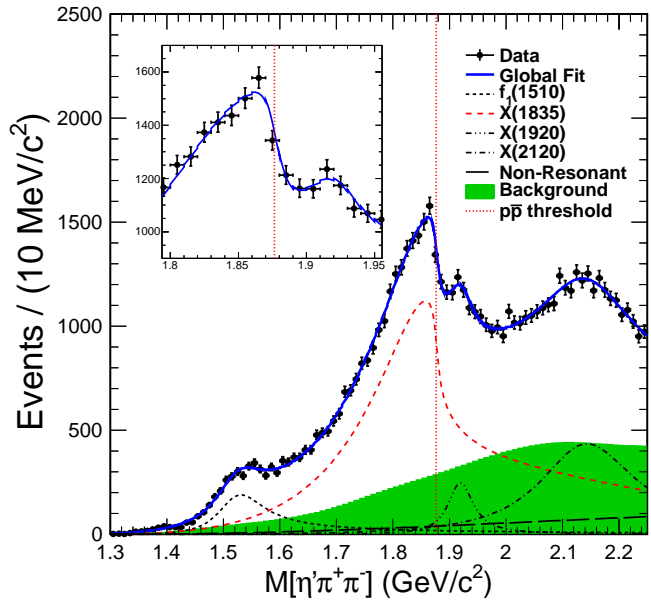}
  \caption{Fit results of using Flatt\'{e} formula.
           The dashed dotted vertical line shows the position of $\ppbar$ mass threshold,
           the dots with error bars are data, the solid curves are total fit results, the dashed curves are the state around $1.85~\gev$,
           the short-dashed curves are the $f_{1}(1510)$, the dash-dotted curves are the $X(2120)$,
           the dash-dot-dot-dotted curves are $X(1920)$,
           and the long-dashed curves are non-resonant $\etap\pip\pim$ fit results;
           the shaded histograms are background events.
           The inset shows the data and the global fit between $1.8~\gev$ and $1.95~\gev$.}
  \label{fig_fit_flatte}
\end{figure}

The fit results for this model are shown in Fig.~\ref{fig_fit_flatte}.
The Flatt\`{e} model fit has a $\logl=630549.5$ that is improved
over the simple Breit-Wigner one by 46, so the significance of $\gpp$ being non-zero is $9.6\sigma$.
In the fit, an additional Breit-Wigner resonance (denoted as ``$X(1920)$'' in
Fig.~\ref{fig_fit_flatte}) is needed with a mass of
$1918.6\pm3.0~\mev$ and width of
$50.6\pm20.9~\mev$; the statistical significance of this
peak is $5.7\sigma$.
In the simple Breit-Wigner fit, the significance of $X(1920)$ is negligible.
The fit yields $\mathcal{M} =
1638.0\pm121.9~\mev$, $g^{2}_{0}
=93.7\pm35.4~(\gev)^{2}$, $\gpp=2.31\pm0.37$, a
product branching fraction of $\mathcal{B}$($\jpsi\to\gamma X$)$\cdot\mathcal{B}$($X\to\etap\pip\pim)=(3.93\pm0.38)\times10^{-4}$.
The value of $\gpp$ implies that the couplings between the state
around $1.85~\gev$ and the $\ppbar$ final states is very large.
Following the definitions given in Ref.~\cite{pole},
the pole position is determined by requiring the denominator in
Eq.~\ref{eq_flatte} to be zero.
The pole nearest to the $\ppbar$
mass threshold is found to be
$M_\text{pole}=1909.5\pm15.9~\mev$ and
$\Gamma_\text{pole}=273.5\pm21.4~\mev$. Taking the systematic
uncertainties (see below) into account,
the significance of $\gpp$ being non-zero is larger than $7\sigma$.


In the second model, we assume the existence of a narrow resonance
near the $\ppbar$ threshold and that the interference between this
resonance and the $X(1835)$ produces the line shape distortion.
Here we denote this narrow resonance as ``$X(1870)$.''
For this case we represent the line shape in the vicinity on 1835~MeV by the square of $T$, where
\eq{
  T=\left(\frac{\sqrt{\rho_\text{out}}}{M_{1}^{2}-s-iM_{1}\Gamma_{1}}+
\frac{\beta e^{i\theta}\sqrt{\rho_\text{out}}}{M_{2}^{2}-s-iM_{2}\Gamma_{2}}\right).
  \label{eq_interference}
}
Here, $\rho_\text{out}$ and $s$ have the same meaning as they had in Eq.~\ref{eq_flatte};
$M_{1}$, $\Gamma_{1}$, $M_{2}$ and $\Gamma_{2}$ represent the masses and widths of the $X(1835)$ and $X(1870)$
resonances respectively; and
$\beta$ and $\theta$ are the relative $\etap\pip\pim$ coupling strengths and the phase between the two resonances.

\begin{figure}[htbp]
  \centering
  \includegraphics[width=0.48\textwidth]{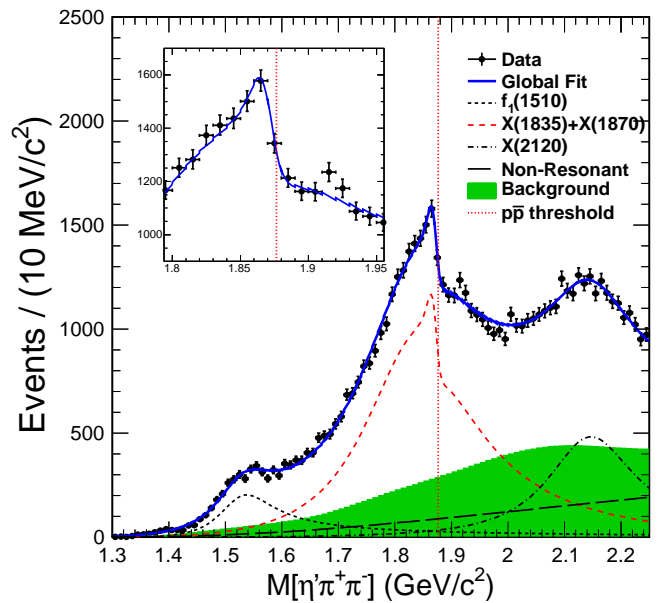}
  \caption{Fit results of using a coherent sum of two Breit-Wigner amplitudes.
           The dashed dotted vertical line shows the position of $\ppbar$ mass threshold,
           the dots with error bars are data, the solid curves are total fit results, the dashed curves are the sum of
           $X(1835)$ and $X(1870)$,
           the short-dashed curves are the $f_{1}(1510)$, the dash-dotted curves are the $X(2120)$,
           the long-dashed curves are non-resonant $\etap\pip\pim$ fit results,
           the shaded histograms are background events.
           The inset shows the data and the global fit between $1.8~\gev$ and $1.95~\gev$.}
  \label{fig_fit_interference}
\end{figure}

The fit results for the second model are shown in Fig.~\ref{fig_fit_interference}.
The $\logl$ of this fit is 630540.3, which is improved by 37 with 4 additional parameters
over that for the fit using one simple Breit-Wigner.
The $X(1835)$ mass is
$1825.3\pm2.4~\mev$ and width is
$245.2\pm13.1~\mev$; The $X(1870)$ mass is
$1870.2\pm2.2~\mev$ and width is $13.0\pm6.1~\mev$, with a statistical significance that is $7.9\sigma$.
It is known that there are two non-trivial solutions in a fit using
a coherent sum of two Breit-Wigner
functions~\cite{multisolution}. In the parameterization of
Eq.~\ref{eq_interference}, the two solutions share the same $M_{1}$,
$\Gamma_{1}$, $M_{2}$ and $\Gamma_{2}$, but have different values of $\beta$
and $\theta$, which means that the only observable difference between the solutions
are branching fractions of the two Breit-Wigner functions. The product branching
fractions  with constructive interference are
 $\mathcal{B}$($\jpsi\to\gamma
X(1835)$)$\cdot\mathcal{B}$($X(1835)\to\etap\pip\pim)$ =
$(3.01\pm0.17)\times10^{-4}$ and
$\mathcal{B}$($\jpsi\to\gamma
X(1870)$)$\cdot\mathcal{B}$($X(1870)\to\etap\pip\pim)$ =
$(2.03\pm0.12)\times10^{-7}$, while the solution with
destructive interference gives $\mathcal{B}$($\jpsi\to\gamma
X(1835)$)$\cdot\mathcal{B}$($X(1835)\to\etap\pip\pim)=(3.72\pm0.21)\times10^{-4}$,
and $\mathcal{B}$($\jpsi\to\gamma
X(1870)$)$\cdot\mathcal{B}$($X(1870)\to\etap\pip\pim)=(1.57\pm0.09)\times10^{-5}$.
In this model, the $X(1920)$ is not included in the fit because its significance is just $3.9\sigma$.
Considering systematic uncertainties (see below),
the significance of $X(1870)$ is larger than $7\sigma$.

The systematic uncertainties come from data-MC differences in the tracking, photon detection and particle identification
efficiencies, the kinematic fit, requirements on the invariant mass distribution of $\gamma\gamma$, signal selection of $\rho^{0}$, $\eta$ and $\etap$,
total number of $\jpsi$ events,
branching fractions for intermediate states decays, fit ranges, background descriptions, mass resolutions, and
intermediate structure of $\pip\pim$.
In the first model, the dominant terms are the fit range, the background description and the intermediate structure of $\pip\pim$.
Considering all systematic uncertainties, the final result is shown in Table~\ref{tab_flatte}.
\begin{table}
  \begin{center}
  \caption{Fit results of using Flatt\'{e} formula.
           The first errors are statistical errors, the second errors are systematic errors;
           the branching ratio is the product of $\mathcal{B}$($\jpsi\to\gamma X)$ and $\mathcal{B}$($X\to\etap\pip\pim)$.}
  \label{tab_flatte}
  \begin{tabular*}{8.3cm}{l p{1.8cm} p{3.8cm}}
    \hline\hline
    \multicolumn{3}{l}{The state around $1.85~\gev$} \\
    \hline
    $\mathcal{M}$ ($\mev$)     & & $1638.0\pm121.9^{+127.8}_{-254.3}$ \\
    $g^{2}_{0}$ ($(\gev)^{2}$) & & $93.7\pm35.4^{+47.6}_{-43.9}$ \\
    $\gpp$                     & & $2.31\pm0.37^{+0.83}_{-0.60}$ \\
    $M_\text{pole}$ ($\mev$)        & & $1909.5\pm15.9^{+9.4}_{-27.5}$ \\
    $\Gamma_\text{pole}$ ($\mev$)   & & $273.5\pm21.4^{+6.1}_{-64.0}$ \\
    Branching Ratio            & & $(3.93\pm0.38^{+0.31}_{-0.84})\times 10^{-4}$ \\
    \hline\hline
  \end{tabular*}
  \end{center}
\end{table}
For the second model, the dominant two systematic sources are the
background description and the intermediate structure of $\pip\pim$.
Considering all systematic uncertainties, the final result is shown in Table~\ref{tab_interference}.
\begin{table}
  \begin{center}
  \caption{Fit results using a coherent sum of two Breit-Wigner amplitudes.
           The first errors are statistical errors, the second errors are systematic errors;
           the branching ratio is the product of $\mathcal{B}$($\jpsi\to\gamma X)$ and $\mathcal{B}$($X\to\etap\pip\pim)$.}
  \label{tab_interference}
  \begin{tabular*}{8.3cm}{l p{3.8cm}}
    \hline\hline
    \multicolumn{2}{l}{$X(1835)$} \\
    \hline
    Mass ($\mev$)                    & $1825.3\pm2.4^{+17.3}_{-2.4}$ \\
    Width ($\mev$)                   & $245.2\pm13.1^{+4.6}_{-9.6}$ \\
    B.R. (constructive interference) & $(3.01\pm0.17^{+0.26}_{-0.28})\times 10^{-4}$ \\
    B.R. (destructive interference)  & $(3.72\pm0.21^{+0.18}_{-0.35})\times 10^{-4}$ \\
    \hline
    \multicolumn{2}{l}{$X(1870)$} \\
    \hline
    Mass ($\mev$)                    & $1870.2\pm2.2^{+2.3}_{-0.7}$ \\
    Width ($\mev$)                   & $13.0\pm6.1^{+2.1}_{-3.8}$ \\
    B.R. (constructive interference) & $(2.03\pm0.12^{+0.43}_{-0.70})\times 10^{-7}$ \\
    B.R. (destructive interference)  & $(1.57\pm0.09^{+0.49}_{-0.86})\times 10^{-5}$ \\
    \hline\hline
  \end{tabular*}
  \end{center}
\end{table}


In summary, the $\jpsi\to\gamma\etap\pip\pim$ process is studied
with $1.09\times10^{9}$ $\jpsi$ events collected at the BESIII
experiment in 2012. We observed a significant distortion of the
$\etap\pip\pim$ line shape near the $\ppbar$ mass threshold that
cannot be accommodated  by an ordinary Breit-Wigner resonance
function. Two typical models for such a line shape are used to fit
the data. The first model assumes the state around $1.85~\gev$
couples with $\ppbar$ and the distortion reflects the opening of the
$\ppbar$ decay channel. The fit result for this model yields a strong
coupling between the broad structure and the $\ppbar$ of
$\gpp=2.31\pm0.37^{+0.83}_{-0.60}$,
with a statistical significance larger than $7\sigma$ for being non-zero.
The pole nearest to the $\ppbar$ mass threshold of this state is located at
$M_\text{pole}=1909.5\pm15.9(\text{stat.})^{+9.4}_{-27.5}(\text{\text{syst.}})~\mev$
and
$\Gamma_\text{pole}=273.5\pm21.4(\text{stat.})^{+6.1}_{-64.0}(\text{syst.})~\mev$.
The second model assumes the distortion reflects interference
between the $X(1835)$ and another resonance with mass close to the
$\ppbar$ mass threshold.  A fit with this model uses a coherent
sum of two interfering Breit-Wigner amplitudes to describe the
$\etap\pip\pim$ mass spectrum around $1.85~\gev$. This fit yields a
narrow resonance below the $\ppbar$ mass threshold with
$M=1870.2\pm2.2(\text{stat.})^{+2.3}_{-0.7}(\text{syst.})~\mev$ and
$\Gamma=13.0\pm6.1(\text{stat.})^{+2.1}_{-3.8}(\text{syst.})~\mev$,
with a statistical significance larger than $7\sigma$.
With current data, both models fit the data well with fit qualities,
and both suggest the existence of a state,
either a broad state with strong couplings to $\ppbar$,
or a narrow state just below the $\ppbar$ mass threshold.
For the broad state above the $\ppbar$ mass
threshold, its strong couplings to $\ppbar$ suggests the existence
of a $\ppbar$ molecule-like state. For the narrow state just below
$\ppbar$ mass threshold, its very narrow width suggests that it be an
unconventional meson, most likely a $\ppbar$ bound state. So
both fits support the existence of a $\ppbar$ molecule-like or bound
state. With current statistics, more sophisticated models such as a mixture of
above two models cannot be ruled out.
In order to elucidate further the nature of the states around $1.85~\gev$,
more data are needed to further study $\jpsi\to\gamma\etap\pip\pim$ process.
Also, line shapes for other decay modes should be studied near the $\ppbar$
mass threshold, including further studies of $\jpsi\to\gamma\ppbar$
and $\jpsi\to\gamma\eta K^{0}_{S}K^{0}_{S}$.

The BESIII collaboration thanks the staff of BEPCII and the IHEP computing center for their strong support.
This work is supported in part by National Key Basic Research Program of China under Contract No. 2015CB856700;
National Natural Science Foundation of China (NSFC) under Contracts Nos. 11235011, 11322544, 11335008, 11425524;
the Chinese Academy of Sciences (CAS) Large-Scale Scientific Facility Program;
the CAS Center for Excellence in Particle Physics (CCEPP); the Collaborative Innovation Center for Particles and Interactions (CICPI);
Joint Large-Scale Scientific Facility Funds of the NSFC and CAS under Contracts Nos. U1232201, U1332201;
CAS under Contracts Nos. KJCX2-YW-N29, KJCX2-YW-N45;
100 Talents Program of CAS;
National 1000 Talents Program of China;
INPAC and Shanghai Key Laboratory for Particle Physics and Cosmology;
Istituto Nazionale di Fisica Nucleare, Italy;
Joint Large-Scale Scientific Facility Funds of the NSFC and CAS	under Contract No. U1532257;
Joint Large-Scale Scientific Facility Funds of the NSFC and CAS under Contract No. U1532258;
Koninklijke Nederlandse Akademie van Wetenschappen (KNAW) under Contract No. 530-4CDP03;
Ministry of Development of Turkey under Contract No. DPT2006K-120470;
The Swedish Resarch Council;
U. S. Department of Energy under Contracts Nos. DE-FG02-05ER41374, DE-SC-0010504, DE-SC0012069, DESC0010118;
U.S. National Science Foundation; University of Groningen (RuG) and the Helmholtzzentrum fuer Schwerionenforschung GmbH (GSI), Darmstadt;
WCU Program of National Research Foundation of Korea under Contract No. R32-2008-000-10155-0.

\bibliographystyle{unsrt}

\end{document}